\documentclass[preprint,superscriptaddress,amsmath,amssymb,prd,aps,showpacs,floatfix,nofootinbib]{revtex4-1}
\usepackage{graphicx}
\usepackage{dcolumn}
\usepackage{bm}
\usepackage{mathrsfs}
\usepackage{hyperref}
\usepackage{amsmath}
\usepackage{amssymb}
\usepackage{bm}
\usepackage{color}
\usepackage{float}
\usepackage{dcolumn}
\usepackage{multirow}
\usepackage{changepage}
\usepackage{breakurl}
\hypersetup{pdftex,colorlinks=true,linkcolor=blue,citecolor=red,menucolor=black,urlcolor=blue,filecolor=blue}

\begin{document}
\title{$\Xi_{bb}$ and $\Omega_{bbb}$ molecular states}
\date{\today}
\author{J.~M.~Dias}
\email{isengardjor@gmail.com}
\affiliation{Department of Physics, Guangxi Normal University, Guilin 541004, China}
\affiliation{Instituto de F{\'i}sica, Universidade de S{\~a}o Paulo, Rua do Mat{\~a}o 1371. Butant{\~a}, CEP 05508-090, S{\~a}o Paulo, S{\~a}o Paulo, Brazil}

\author{Qi-Xin~Yu}
\email{yuqx@mail.bnu.edu.cn}
\affiliation{Department of Physics, Guangxi Normal University, Guilin 541004, China}
\affiliation{Institute for Experimental Physics, Department of Physics, University of Hamburg, Luruper Chaussee 149, D-22761 Hamburg, Germany}

\author{Wei-Hong~Liang}
\email{liangwh@gxnu.edu.cn}
\affiliation{Department of Physics, Guangxi Normal University, Guilin 541004, China}
\affiliation{Guangxi Key Laboratory of Nuclear Physics and Technology, Guangxi Normal University, Guilin 541004, China}

\author{Zhi-Feng~Sun}
\affiliation{School of Physical Science and Technology, Lanzhou University, Lanzhou 730000, China}

\author{Ju-Jun~Xie}
\affiliation{Institute of Modern Physics, Chinese Academy of Sciences, Lanzhou 730000, China}
\affiliation{School of Physics and Microelectronics, Zhengzhou University, Zhengzhou, Henan 450001, China}
\affiliation{School of Nuclear Science and Technology, University of Chinese Academy of Sciences, Beijing 100049, China}

\author{E.~Oset}
\email{oset@ific.uv.es}
\affiliation{Department of Physics, Guangxi Normal University, Guilin 541004, China}
\affiliation{Departamento de F\'{\i}sica Te\'orica and IFIC, Centro Mixto Universidad de
Valencia-CSIC Institutos de Investigaci\'on de Paterna, Aptdo.22085,
46071 Valencia, Spain}

\begin{abstract}
Using the vector-exchange interaction in the local hidden gauge approach, which in the light quark sector generates the chiral Lagrangian,
and has produced realistic results for $\Omega_c, \Xi_c, \Xi_b$ and the hidden charm pentaquark states,
we study the meson-baryon interactions in the coupled channels that lead to the $\Xi_{bb}$ and $\Omega_{bbb}$ excited states of the molecular type.
We obtain seven states of the $\Xi_{bb}$ type with energies between $10408$ and $10869$ MeV and one $\Omega_{bbb}$ state at $15212$ MeV.
\end{abstract}

\maketitle


\section{Introduction}
\label{sec:intro}
Doubly- and triply-heavy baryons have attracted continuous theoretical attention \cite{Olsen,Karliner,slzhu}
which has been intensified with the recent finding of the $\Xi_{cc}^{++}$ state by LHCb \cite{XiLHCb}.
$\Xi_{bb}$ and $\Omega_{bbb}$ states have not yet been found,
but are likely to be observed in the future by the LHCb or Belle II collaborations \footnote{The effective efficiency for reconstruction of beauty hadrons is small, which makes the reconstruction of two beauty hadrons difficult. Yet, searches for such states are going on at LHCb, and such states should become more accessible in future updates of present facilities \cite{Belyaev}.}.
Thus, it is appropriate to make theoretical predictions before the experiments are performed.
Concerning the $\Xi_{bb}$ and $\Omega_{bbb}$ states, most of the theoretical work concentrated on quark model calculations with three quarks.
Pioneering work in this field was presented in Ref.~\cite{Rujula}.
A reference work on doubly- and triply-heavy baryons is the one of Ref.~\cite{Brac}.
More recently, there has been a theoretical revival stimulated by the new experimental findings and one finds studies of doubly-heavy baryons in Refs.~\cite{Cohen,qixin,Juan,xzWeng,Shah,Refd66,Roberts:2007ni,Refa37,Refd90,Qifang,
Slzhu2017,lumeng,Aliev,KarlinerPRD2018},
and of triply-heavy baryons in Refs.~\cite{Refd91,Refd95,RefA53,Segovia,Shah2018,xzWeng,Radin,Azizi},
among others.
Lattice QCD has also been used to make predictions of these states \cite{Refd85,d82,d90}.
Calculations with pentaquark configurations are available in Refs.~\cite{slZhupen,liliu}.
Suggestions on how to observe these states by looking at weak decay products have been made in Refs.~\cite{weiwang,weiwang2},
and using the $e^+e^-$ colliders in Ref.~\cite{xuchang}.
Yet, molecular states of this type based on the meson-baryon interaction have not been investigated so far,
and this is the purpose of the present work.

Molecular states bound by the meson-baryon strong interaction in coupled channels are peculiar.
While there can be states bound by several tens of MeV,
there are others which are very close to the threshold of meson-baryon channels.
Let's assume, to begin with, that we have just one meson-baryon channel bound by a small binding energy $B$.
The coupling of this state to the meson-baryon component, $g$, is such that $-g^2 \,\frac{\partial G}{\partial E}\Big|_{E_B}=1$ ($E_B$ is the energy of the bound state),
where $G$ is the meson-baryon loop function such that the scattering matrix is given by $T= V+VGT$.
This function has a cusp at the threshold of the meson-baryon channel,
such that its derivative to the left is infinite at the threshold.
Thus, $g^2\to 0$ as the binding $B$ goes to zero \cite{danijuan}.
This can be derived from another perspective and is known as the Weinberg compositeness condition \cite{Weinberg,Kalash}, with $g^2 \sim \sqrt{B}$.
What is less known is that when one has coupled channels and if the bound state is close to the threshold of one of the coupled channels, then
the couplings of the bound state to all channels reduce to zero \cite{juantoki,danijuan}.
As a consequence, the decay widths for the given channels, proportional to $g_i^2$,
go to zero and one obtains automatically very narrow widths.
This property, so naturally obtained with molecular states, is a source of permanent problems in the three-quark or tight pentaquark models of these states \cite{Pilloni}.

A clear situation favoring molecular states is the recent finding of three narrow pentaquark states by the LHCb collaboration
close to the $\Sigma_c \bar D, \Sigma_c \bar D^*$ thresholds \cite{LHCbpenta},
which have been interpreted in a large number of papers as molecular states \cite{Many16,Many17,Many18,Many19,Many21,Many22,Many23,Many24,Many25,Many26,Many27,Many28,Many29,Many30,Many31}.
This is also the case in Ref.~\cite{Xiao}, where the earlier predictions made in Ref.~\cite{WuRequel} were updated.
The same molecular model has been successful in predicting three of the five narrow $\Omega_c$ states found by LHCb \cite{LHCbBom}
in Refs.~\cite{Montana,ViniciusOm,paraom}, of some $\Xi_c$ states reported in PDG \cite{pdg},
and of $\Xi_b (6227)$ observed by the LHCb collaboration \cite{LHCbxib} in Ref.~\cite{Qixincb}.
Predictions of $\Xi_{bc}$ states that have not yet observed were made in Ref.~\cite{QixinLiang}.
We should pointed out that the molecular picture is not the only one which claims to reproduce these states, and a variety of other models have been suggested.
Abundant information can be found in a series of review papers \cite{Olsen,Karliner,slzhu,Guo37,Guo38,Guo39,Guo40,Guo41,Guo42,Guo43,Guo44,Guo45,
Guo47,Guo48,Guo51}.
In this sense, making predictions with different models prior to experiments
is useful to gain a better understanding of the nature of these states.
Our work is written in this perspective.

We use here the same source of interaction that has been tested successfully in other cases and make predictions for the $\Xi_{bb}$ and $\Omega_{bbb}$ states.

\section{Formalism}
\label{sec:form}
\begin{figure}[b]
\begin{center}
\includegraphics[scale=0.65]{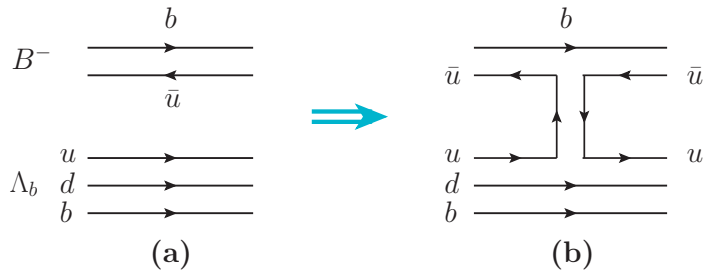}
\end{center}
\vspace{-0.7cm}
\caption{(a) Quark representation of $B^- \Lambda_b$; (b) Meson exchange mechanism for the $B^- \Lambda_b \to B^- \Lambda_b$ transition.}
\label{Fig:1}
\end{figure}

In order to understand the classification of the meson-baryon states considered here,
it is convenient to begin with the interaction we use.
Let us look, as an example, at the $B^- \Lambda_b \to B^- \Lambda_b$ transition
shown in Fig.~\ref{Fig:1}.
By means of the mechanism of Fig.~\ref{Fig:1}(b), one can exchange a $u\bar u$ state between $B^-$ and $\Lambda_b$.
This could physically correspond to a $\pi$, $\rho$ or $\omega$ meson.
One can equally exchange a $b\bar b$ pair which could correspond to $\eta_b$ or $\Upsilon$,
but we can anticipate that $\eta_b$ or $\Upsilon$ exchange, corresponding to a meson propagator, would be very much suppressed
because of the large mass of the $b\bar b$ state compared to $u\bar u$.

We next consider the case where we have $K^-$ instead of $B^-$ in Fig.~\ref{Fig:1},
where the chiral Lagrangians can be used to obtain the strength of the exchange mechanism.
We recall the observation in Ref.~\cite{deRafael} that the chiral Lagrangians can be obtained from the local hidden gauge Lagrangians
which rely on the exchange of vector mesons \cite{hidden1,hidden2,hidden4,hideko}.
In the $K^- \Lambda_b \to K^- \Lambda_b$ interaction, we have the $u\bar u$ exchange and the $s$ quark as a spectator,
the same as in the diagram Fig.~\ref{Fig:1}(b) where the $b$ quark is a spectator.
We can make a mapping from the $K^- \Lambda_b \to K^- \Lambda_b$ interaction to the $B^- \Lambda_b \to B^- \Lambda_b$ interaction at the quark level,
taking into account that when writing the $S$ matrix at the meson level
the normalization factors of the meson fields $\frac{1}{\sqrt{2 E_K}}$, $\frac{1}{\sqrt{2 E_B}}$ are different.
These considerations were made in Ref.~\cite{XiaoLiang}.

In the evaluation of the $B^- \Lambda_b \to B^- \Lambda_b$ transition in Fig.~\ref{Fig:1}(b), instead of the Lagrangians one can use the operators at the quark level,
both in the upper vertex $BBV$ \cite{Xibc68} and in the lower vertex $\Lambda_b \Lambda_b V$, with $V$ the exchanged vector meson \cite{ViniciusOm},
to get the same result. For practical reasons, we use the Lagrangian for the upper vertex
\begin{equation}\label{eq:Lagrangian}
\mathcal L=-ig\, \langle\left[P,\partial_\mu P\right]V^\mu\rangle,
\end{equation}
where $\langle\cdots\rangle$ stands for the matrix trace, $g=\frac{M_V}{2f_\pi}$ ($M_V \sim 800\,\rm MeV$ is the vector mass, $f_\pi=93\,\rm MeV$),
and $P, V$ are the $q\bar q$ matrices written in terms of pseudoscalar or vector mesons,
with quarks $u,d,s,b$.
Hence
\begin{equation}
\label{eq:matP}
P = \begin{pmatrix}
\frac{1}{\sqrt{2}}\pi^0 + \frac{1}{\sqrt{3}} \eta + \frac{1}{\sqrt{6}}\eta' & \pi^+ & K^+ & B^+ \\
 \pi^- & -\frac{1}{\sqrt{2}}\pi^0 + \frac{1}{\sqrt{3}} \eta + \frac{1}{\sqrt{6}}\eta' & K^0 & B^0 \\
 K^- & \bar{K}^0 & -\frac{1}{\sqrt{3}} \eta + \sqrt{\frac{2}{3}}\eta' & B_s^0 \\
B^-  & \bar B^0 & \bar B_s^0 & \eta_b
\end{pmatrix},
\end{equation}
\begin{equation}
\label{eq:matV}
V = \begin{pmatrix}
 \frac{1}{\sqrt{2}}\rho^0 + \frac{1}{\sqrt{2}} \omega & \rho^+ & K^{* +} & B^{* +} \\
 \rho^- & -\frac{1}{\sqrt{2}}\rho^0 + \frac{1}{\sqrt{2}} \omega  & K^{* 0} & B^{* 0} \\
 K^{* -} & \bar{K}^{* 0}  & \phi & B_s^{* 0} \\
 B^{* -} & \bar B^{* 0} & \bar B_s^{* 0} & \Upsilon
\end{pmatrix},
\end{equation}
where we use the $\eta$-$\eta'$ mixing of Ref.~\cite{Bramon}.
The lower vertex is of the type $V_\nu \gamma^\nu$,
and we make the approximation that the momenta of the particles are small compared to their masses
and $\gamma^\nu \to \gamma^0 \equiv 1$, rendering the interaction spin independent.
This means that after contraction of $V^\mu V^\nu$, only the $\partial_0 $ component of $\partial_\mu $ in  Eq.~\eqref{eq:Lagrangian} is operative.
The lower vertex is still evaluated at the quark level and the Lagrangian is trivial in terms of the operators,
\begin{equation}
\mathcal L\rightarrow \left\{\begin{matrix}&\displaystyle\frac{g}{\sqrt2}(u\bar u-d\bar d),\,\,\text{for}\,\,\rho^0\\[0.5cm]
                                      &\displaystyle\frac{g}{\sqrt2}(u\bar u+d\bar d),\,\,\text{for}\,\,\omega\end{matrix}\right.
\end{equation}
which has to be sandwiched between the baryon wave functions.
The next step to complete the program is to write the wave functions,
and here we divert from using SU(4) or other extensions of SU(5), because the heavy quarks are not identical particles to the $u,d,s$ quarks.
We single out the heavy quark and impose the flavor-spin symmetry on the light quarks.
If instead we have two or three $b$ quarks,
then we impose the flavor-spin symmetry on the $b$ quarks.
Once this is clarified, we have the follwing meson-baryon states with two $b$ quarks:
\begin{itemize}
  \item [1)] $bb$ in the baryon,
  \begin{equation*}
    \Xi_{bb}, ~ \Omega_{bb},
  \end{equation*}
  and the meson-baryon states are
      \begin{equation}
      \pi \,\Xi_{bb},\quad \eta \,\Xi_{bb},\quad K \,\Omega_{bb}.
      \end{equation}
  Since we have two identical $b$ quarks, the spin wave function has to be symmetric in these quarks.
  We take them as number $1$ and $2$,
  and thus we must use the mixed symmetric spin wave function $\chi_{MS}$ for the first two quarks.
  \item [2)] One $b$ quark in the baryon and one in the meson. The meson-baryon states are
  \begin{equation}
  \bar B \,\Lambda_b,\quad \bar B \,\Sigma_b,\quad \bar B_s \,\Xi_b,\quad \bar B_s \,\Xi_b^\prime.
  \end{equation}
  In this case the flavor-spin symmetry is imposed on the second and third (light) quarks.
  The baryon states are classified as shown in Table~\ref{tab:baryonwave}.
  \end{itemize}
\begin{table}[t]
\renewcommand\arraystretch{1.0}
\centering
\caption{\vadjust{\vspace{-0pt}}Wave functions for baryons with $J^P=\frac{1}{2}^+$ and $I=0,\frac{1}{2},1$.
$MS$ and $MA$ stand for mixed symmetric and mixed antisymmetric, respectively.}\label{tab:baryonwave}
\begin{tabular*}{0.60\textwidth}{@{\extracolsep{\fill}}cccc}
\hline
\hline
States         & $I,J$     & Flavor                       & Spin   \\
\hline
$\Xi^0_{bb}$& $\frac{1}{2},\, \frac{1}{2}$   & $bbu$                        & $\chi_{MS}(12)$\\
$\Omega^-_{bb}$  & $0,\, \frac{1}{2}$   & $bbs$    & $\chi_{MS}(12)$\\
$\Lambda_b^0$   & $0, \, \frac{1}{2}$   & $b\frac{1}{\sqrt2}(ud-du)$        & $\chi_{MA}(23)$\\
$\Sigma_b^0$      & $1, \, \frac{1}{2}$ & $b\frac{1}{\sqrt2}(ud+du)$    & $\chi_{MS}(23)$\\
$\Xi_b^0$        &$\frac{1}{2},\, \frac{1}{2}$& $b\frac{1}{\sqrt2}(us-su)$    & $\chi_{MA}(23)$\\
$\Xi_b^{'0}$  & $\frac{1}{2},\, \frac{1}{2}$   & $b\frac{1}{\sqrt2}(us+su)$    & $\chi_{MS}(23)$\\
\hline
\hline
\end{tabular*}
\end{table}

We need $\chi_{MS}(12)$, $\chi_{MS}(23)$ and $\chi_{MA}(23)$, which are given in Ref.~\cite{Close} for $s_3 =\frac{1}{2}$,
\begin{align}\label{eq:chiMSMA}
  \chi_{MS}(12) &= \frac{1}{\sqrt{6}} (\uparrow \downarrow \uparrow +\downarrow\uparrow\uparrow-2\uparrow\uparrow\downarrow), \\
  \chi_{MS}(23) &= \frac{1}{\sqrt{6}} (\uparrow\uparrow\downarrow+\uparrow\downarrow\uparrow-2\downarrow\uparrow\uparrow), \\
  \chi_{MA}(23) &= \frac{1}{\sqrt{2}} (\uparrow\uparrow\downarrow-\uparrow\downarrow\uparrow).
\end{align}
Note that with our spin independent interaction, $\Xi_{bb}$ and $\Omega_{bb}$ can have spin overlap with the other baryon components of Table~\ref{tab:baryonwave},
since
\begin{align}
  \langle \chi_{MS}(12)| \chi_{MS}(23) \rangle &= -\frac{1}{2}, \\
  \langle \chi_{MS}(12)| \chi_{MA}(23) \rangle & =-\frac{\sqrt{3}}{2}.
\end{align}
We also consider vector-baryon states and pseudoscalar combinations with baryons with $J^P=\frac{3}{2}^+$,
$\Xi^*_{bb}$, $\Omega^*_{bb}$, $\Sigma_b^*$, $\Xi^*_b$, shown in Table~\ref{tab:baryonwave2}.
They all have the full symmetric spin wave function $\chi_S$,
\begin{equation}\label{eq:chiS}
  \chi_S(s_3=1)=\uparrow\uparrow\uparrow.
\end{equation}

\begin{table}[t]
\renewcommand\arraystretch{1.0}
\centering
\caption{\vadjust{\vspace{-8pt}}Wave functions for baryons with $J^P=\frac{3}{2}^+$ and $I=0,\frac{1}{2},1$.
$S$ in $\chi_S$ stands for full symmetric.}\label{tab:baryonwave2}
\begin{tabular*}{0.60\textwidth}{@{\extracolsep{\fill}}cccc}
\hline
\hline
States         & $I,J$     & Flavor                       & Spin   \\
\hline
$\Xi_{bb}^{\ast0}$& $\frac{1}{2}, \, \frac{3}{2}$   & $bbu$                  & $\chi_{S}$\\
$\Omega_{bb}^{\ast-}$   & $0,\, \frac{3}{2}$   & $bbs$                  & $\chi_{S}$\\
$\Sigma_b^{\ast0}$      & $1,\, \frac{3}{2}$ & $b\frac{1}{\sqrt2}(ud+du)$    & $\chi_{S}$\\
$\Xi_b^{\ast0}$& $\frac{1}{2},\, \frac{3}{2}$   & $b\frac{1}{\sqrt2}(us+su)$         & $\chi_{S}$\\
\hline
\hline
\end{tabular*}
\end{table}

The combination of vector-baryon($\frac{3}{2}^+$) gives rise to states in a region difficult to identify experimentally \cite{Xiao} and we do not study them.

In the case of vector-baryon interaction,
the upper vertex is evaluated using Eq.~\eqref{eq:Lagrangian} substituting $\left[P,\partial_\mu P\right]$ with $\left[V_\nu,\partial_\mu V^\nu\right]$.
In the limit of small momenta,
$V^\mu$ of Eq.~\eqref{eq:Lagrangian} is the exchanged vector.
Hence, the interaction is calculated in the same way as for pseudoscalars except that there is an extra $\vec \epsilon \cdot \vec \epsilon\,'$ factor
which gives the spin independence of the upper vertex \cite{ViniciusOm}.
Therefore, the interaction is spin independent.
This feature allows us to classify the channels into different blocks:
\begin{itemize}
  \item [a)] $\pi \,\Xi_{bb}$, $\eta \,\Xi_{bb}$ and $K \,\Omega_{bb}$ with $\chi_{MS}(12)$, $\bar B \,\Lambda_b$ and $\bar B_s \,\Xi_b$ with $\chi_{MA}(23)$;
  \item [b)] $\pi \,\Xi_{bb}$, $\eta \,\Xi_{bb}$ and $K \,\Omega_{bb}$ with $\chi_{MS}(12)$, $\bar B \,\Sigma_b$ and $\bar B_s \,\Xi'_b$ with $\chi_{MS}(23)$;
  \item [c)] $\rho \,\Xi_{bb}$, $\omega\,\Xi_{bb}$, $\phi \,\Xi_{bb}$ and $K^\ast \,\Omega_{bb}$ with $\chi_{MS}(12)$, $\bar B^\ast \,\Lambda_{b}$ and $\bar B_s^\ast \,\Xi_b$ with $\chi_{MA}(23)$;
  \item [d)] $\rho \,\Xi_{bb}$, $\omega \,\Xi_{bb}$, $\phi \,\Xi_{bb}$ and $K^\ast \,\Omega_{bb}$ with $\chi_{MS}(12)$, $\bar B^\ast \,\Sigma_b$ and $\bar B_s^\ast \,\Xi_b^\prime$ with $\chi_{MS}(23)$;
  \item [e)] $\pi \,\Xi^*_{bb}$, $\eta \,\Xi^*_{bb}$, $K \,\Omega^*_{bb}$, $\bar B \,\Sigma^*_b$ and $\bar B_s \,\Xi^*_b$ with all states in $\chi_{S}$.
\end{itemize}

Taking into account our isospin phase convention $(-\pi^+, \pi^0, \pi^-)$, $(B^+, B^0)$, $(\bar B^0, -B^-)$, $(K^+, K^0)$ and $(\bar K^0, -K^-)$,
we can construct the isospin wave function for the blocks to have isospin $I=\frac{1}{2}$ for the global ``$\Xi_{bb}$" states.,
Using the vector-exchange interaction discussed above,
we obtain a potential $V_{ij}$ for the $i \to j$ transition of the type
\begin{equation}\label{eq:kernelnr}
V_{ij} = D_{ij}\, \frac{1}{4 f_{\pi}^2}\, (k^0+k'^0),
\end{equation}
where, $k^0, k'^0$ are the meson energies in channel $i$ and channel $j$, respectively,
and $D_{ij}$ are the coefficients which are given in the tables below.

Note that since $\chi_{MS}(12)$ in the $\Xi_{bb}$, $\Omega_{bb}$ states overlaps with $\chi_{MS}(23)$ and $\chi_{MA}(23)$, blocks a) and b) can mix and have to be put together.
The same can be said for blocks c) and d), which also have to be put together.
We then obtain the $D_{ij}$ coefficients shown in Tables \ref{tab:4}, \ref{tab:6}, \ref{tab:8} (note that we changed the order of baryon-meson in the tables,
which must be taken into account when constructing the isospin wave functions).
In Tables \ref{tab:3}, \ref{tab:5}, \ref{tab:7} we show the thresholds of the channels considered.
\begin{table*}[t]
\renewcommand\arraystretch{1.0}
\begin{adjustwidth}{-0.00\textwidth}{-0.00\textwidth}
\caption{\vadjust{\vspace{-0pt}}Pseudoscalar-baryon($\frac{1}{2}^+$) (PB) channels considered for the sector with $J^P=\frac{1}{2}^-$.}\label{tab:3}
\begin{tabular*}{0.90\textwidth}{@{\extracolsep{\fill}}c|ccccccc}
\hline
\hline
\textbf{Channel} & $\Xi_{bb}\,\pi$ & $\Xi_{bb} \,\eta$ & $\Omega_{bb} \,K$ & $\Lambda_b \,\bar B$ & $\Sigma_b \,\bar B$ & $\Xi_b \,\bar B_s$ & $\Xi_b^\prime \,\bar B_s$  \\
\hline
\textbf{Threshold (MeV)} & $10335$ & $10745$ & $10756$ & $10899$ & $11092$ & $11160$ & $11302$\\
\hline
\hline
\end{tabular*}
\end{adjustwidth}
\end{table*}
The masses which are not tabulated in PDG \cite{pdg} are taken from Refs.~\cite{Juaneli,Roberts:2007ni}.
In the tables there are terms which go with parameter $\lambda$.
They correspond to transitions that require $B^*$ exchange.
Because of the large mass of $B^*$ compared to the light vectors, these terms are very much suppressed.
With the same considerations as in Ref.~\cite{ViniciusOm}, we estimate $\lambda$ as
\begin{eqnarray}
\lambda = \frac{-m^2_{V}}{(m_B - m_\eta)^2 - m^2_{B^*}}\approx 0.1 \, .
\end{eqnarray}
We note that when the light vector mesons are exchanged,
the heavy quarks are spectators, and hence these terms automatically fulfill the rules of heavy quark symmetry.
The exchange of $B^*$ makes the $b$ quark active.
This term goes barely as ${\mathcal O}(\frac{1}{m_Q})$ (with $m_Q$ the heavy quark mass)
and is not subject to the heavy quark spin symmetry rules.
Note that these terms are very small in our approach, as expected.

\begin{table*}[t]
\renewcommand\arraystretch{1.0}
\begin{adjustwidth}{-0.00\textwidth}{-0.0\textwidth}
\caption{\vadjust{\vspace{-0pt}}$D_{ij}$ coefficients for the PB sector with $J^P=\frac{1}{2}^-$.}\label{tab:4}
\begin{tabular*}{0.80\textwidth}{@{\extracolsep{\fill}}c|ccccccc}
\hline
\hline
\textbf{$J^P=\frac{1}{2}^-$} & $\Xi_{bb} \,\pi$ & $\Xi_{bb} \,\eta$ & $\Omega_{bb} \,K$ & $\Lambda_b \,\bar B$ & $\Sigma_b \,\bar B$ & $\Xi_b \,\bar B_s$ & $\Xi_b^\prime \,\bar B_s$\\
\hline
$\Xi_{bb} \,\pi$ & $-2$ & $0$ & $\sqrt\frac{3}{2}$ & $\frac{3}{4}\lambda$ & $-\frac{1}{4}\lambda$ & $0$ & $0$ \\
\hline
$\Xi_{bb} \,\eta$ &  & $0$ & $-\frac{2}{\sqrt 3}$ & $\frac{1}{2\sqrt 2}\lambda$ & $\frac{1}{2\sqrt 2}\lambda$ & $-\frac{1}{2\sqrt 2}\lambda$ & $\frac{1}{2\sqrt 6}\lambda$ \\
\hline
$\Omega_{bb} \,K$ &  &  & $-1$ & $0$ & $0$ & $-\sqrt\frac{3}{8}\lambda$ & $-\frac{1}{2\sqrt 2}\lambda$\\
\hline
$\Lambda_b \,\bar B$ &  &  &  & $-1$ & $0$ & $-1$ & $0$ \\
\hline
$\Sigma_b \,\bar B$ &  &  &  &  & $-3$ & $0$ & $\sqrt 3$ \\
\hline
$\Xi_b  \,\bar B_s$ &  &  &  &  &  & $-1$ & $0$ \\
\hline
$\Xi_b^\prime \,\bar B_s$ &  &  &  &  &  &  & $-1$ \\
\hline
\hline
\end{tabular*}
\end{adjustwidth}
\end{table*}

\begin{table*}[h]
\renewcommand\arraystretch{1.0}
\centering
\caption{\vadjust{\vspace{-0pt}}Vector-baryon($\frac{1}{2}^+$) (VB) channels considered for the sector with $J^P=\frac{1}{2}^-,\frac{3}{2}^-$.}\label{tab:5}
\begin{tabular*}{1.00\textwidth}{@{\extracolsep{\fill}}c|ccccccccc}
\hline
\hline
\textbf{Channel} & $\Lambda_b\bar B^\ast$ & $\Xi_{bb} \rho$ & $\Xi_{bb}\omega$ & $\Sigma_b \bar B^\ast$ & $\Omega_{bb}K^\ast$ & $\Xi_b \bar B_s^\ast$ & $\Xi_{bb}\phi$ & $\Xi_b^\prime\bar B_s^\ast$ \\
\hline
\textbf{Threshold (MeV)} & $10945$ & $10972$ & $10980$ & $11138$ & $11156$ & $11208$ & $11216$ & $11350$ \\
\hline
\hline
\end{tabular*}
\end{table*}

\begin{table*}[h]
\renewcommand\arraystretch{1.0}
\begin{adjustwidth}{0\textwidth}{0\textwidth}
\caption{\vadjust{\vspace{-0pt}}$D_{ij}$ coefficients for the VB sector with $J^P=\frac{1}{2}^-,\frac{3}{2}^-$.}\label{tab:6}
\begin{tabular*}{0.9\textwidth}{@{\extracolsep{\fill}}c|ccccccccc}
\hline
\hline
\textbf{$J^P=\frac{1}{2}^-,\,\frac{3}{2}^-$} & $\Lambda_b \,\bar B^\ast$ & $\Xi_{bb} \,\rho$ & $\Xi_{bb} \,\omega$ & $\Sigma_b \,\bar B^\ast$ & $\Omega_{bb} \,K^\ast$ & $\Xi_b \,\bar B_s^\ast$ & $\Xi_{bb}\,\phi$ & $\Xi_b^\prime \,\bar B_s^\ast$ \\
\hline
$\Lambda_b \,\bar B^\ast$ & $-1$ & $\frac{3}{4}\lambda$ & $\frac{\sqrt 3}{4}\lambda$ & 0 & 0 & $-1$ & 0 & 0 \\
\hline
$\Xi_{bb} \,\rho$ &  & $-2$ & 0 & $-\frac{1}{4}\lambda$ & $\sqrt\frac{3}{2}$ & 0 & 0 & 0 \\
\hline
$\Xi_{bb} \,\omega$ &  &  & 0 & $\frac{\sqrt 3}{4}\lambda$ & $-\frac{1}{\sqrt 2}$ & 0 & 0 & 0 \\
\hline
$\Sigma_b \,\bar B^\ast$ &  &  &  & $-3$ & 0 & 0 & 0 & $\sqrt 3$ \\
\hline
$\Omega_{bb} \,K^\ast$ &  &  &  &  & $-1$ & $\frac{1}{\sqrt 2}\lambda$ & 1 & $-\frac{1}{2\sqrt 2}\lambda$ \\
\hline
$\Xi_b \,\bar B_s^\ast$ &  &  &  &  &  & $-1$ & $\sqrt\frac{3}{8}\lambda$ & 0 \\
\hline
$\Xi_{bb}\,\phi$ &  &  &  &  &  &  & 0 & $-\frac{1}{2\sqrt 2}\lambda$ \\
\hline
$\Xi_b^\prime\bar B_s^\ast$ &  &  &  &  &  &  &  & $-1$ \\
\hline
\hline
\end{tabular*}
\end{adjustwidth}
\end{table*}

\begin{table*}[t]
\renewcommand\arraystretch{1.0}
\centering
\caption{\vadjust{\vspace{-0pt}}Pseudoscalar-baryon($\frac{3}{2}^+$) (PB) channels considered for the sector with $J^P=\frac{3}{2}^-$.}\label{tab:7}
\begin{tabular*}{0.70\textwidth}{@{\extracolsep{\fill}}c|cccccc}
\hline
\hline
\textbf{Channel} & $\Xi_{bb}^*\, \pi$ & $\Xi_{bb}^* \,\eta$ & $\Omega_{bb}^* \,K$ & $\Sigma_b^*  \,\bar B$ & $\Xi_b^* \,\bar B_s$  \\
\hline
\textbf{Threshold (MeV)} & $10374$ & $10784$ & $10793$ & $11113$ & $11320$ \\
\hline
\hline
\end{tabular*}
\end{table*}

\begin{table*}[h]
\renewcommand\arraystretch{1.0}
\centering
\caption{\vadjust{\vspace{-0pt}}$D_{ij}$ coefficients for the PB sector with $J^P=\frac{3}{2}^-$.}\label{tab:8}
\begin{tabular*}{0.60\textwidth}{@{\extracolsep{\fill}}c|cccccccc}
\hline
\hline
\textbf{$J^P=\frac{3}{2}^-$} & $\Xi_{bb}^*\pi$ & $\Xi_{bb}^*\eta$ & $\Omega_{bb}^*K$ & $\Sigma_b^* \bar B$ & $\Xi_b^*\bar B_s$   \\
\hline
$\Xi_{bb}^*\pi$ & $-2$ & 0 & $\sqrt\frac{3}{2}$ & $\frac{1}{2}\lambda$ & 0 \\
\hline
$\Xi_{bb}^*\eta$ &  & 0 & $-\frac{2}{\sqrt 3}$ & $-\frac{1}{\sqrt 2}\lambda$ & $-\frac{1}{\sqrt 6}\lambda$ \\
\hline
$\Omega_{bb}^*K$ &  &  & $-1$ & 0 & $\frac{1}{\sqrt 2}\lambda$\\
\hline
$\Sigma_b^* \bar B$ &  &  &  & $-3$ & $\sqrt 3$ \\
\hline
$\Xi_b^*\bar B_s$ &  &  &  &  & $-1$ \\
\hline
\hline
\end{tabular*}
\end{table*}

\begin{table}[h]
\renewcommand\arraystretch{1.0}
\centering
\caption{\vadjust{\vspace{-0pt}}Wave functions of baryons with $J^P=\frac{3}{2}^+$ and $I=0,\frac{1}{2}$.}\label{tab:baryonwave3}
\begin{tabular*}{0.40\textwidth}{@{\extracolsep{\fill}}cccc}
\hline
\hline
States         & $I,J$     & Flavor                       & Spin   \\
\hline
$\Omega^-_{bbb}$& $0, \, \frac{3}{2}$   & $bbb$                        & $\chi_{S}$\\
$\Xi^{*0}_{bb}$  & $\frac{1}{2}, \,\frac{3}{2}$   & $bbu$    & $\chi_{S}$\\
$\Omega^{*-}_{bb}$   & $0, \, \frac{3}{2}$   & $bbs$        & $\chi_{S}$\\
\hline
\hline
\end{tabular*}
\end{table}

\clearpage
We turn now to the $\Omega_{bbb}$ states.
Here, the coupled channels are $\eta \,\Omega_{bbb}, \bar B \,\Xi^*_{bb}, \bar B_s \,\Omega^*_{bb}$.
The baryon states involved are tabulated in Table \ref{tab:baryonwave3}.
The $\eta\, \Omega_{bbb}$, $\bar B \,\Xi^{*}_{bb}$, $\bar B_s \,\Omega^{*}_{bb}$ channels with $J^P=\frac{3}{2}^-$,
do not couple with the ``$\Xi_{bb}$" states since they contain one more $b$ quark.
In Table \ref{tab:10}, we show the threshold masses of the pseudoscalar-baryon channels, and in Table \ref{tab:11} the $D_{ij}$ coefficients.
\begin{table*}[h]
\renewcommand\arraystretch{1.0}
\centering
\caption{\vadjust{\vspace{-0pt}}PB channels considered for the sector with $J^P=\frac{3}{2}^-$ and three $b$ quarks.}\label{tab:10}
\begin{tabular*}{0.50\textwidth}{@{\extracolsep{\fill}}c|cccc}
\hline
\hline
\textbf{Channel} & $\Omega_{bbb}\,\eta$ &$\Xi_{bb}^* \,\bar B$ & $\Omega_{bb}^* \,\bar B_s$  \\
\hline
\textbf{Threshold (MeV)} & $15382$ & $15515$ & $15664$ \\
\hline
\hline
\end{tabular*}
\end{table*}

\begin{table*}[h]
\renewcommand\arraystretch{1.0}
\centering
\caption{\vadjust{\vspace{-0pt}} $D_{ij}$ coefficients for the PB sector with $J^P=\frac{3}{2}^-$ and three $b$ quarks.}\label{tab:11}
\begin{tabular*}{0.40\textwidth}{@{\extracolsep{\fill}}c|ccc}
\hline
\hline
\textbf{$J^P=\frac{3}{2}^-$} & $\Omega_{bbb}\,\eta$ & $\Xi_{bb}^* \,\bar B$ & $\Omega_{bb}^* \,\bar B_s$   \\
\hline
$\Omega_{bbb}\eta$ & 0 & $-\frac{2}{\sqrt 6}\lambda$ & $-\frac{1}{\sqrt 3}\lambda$ \\
\hline
$\Xi_{bb}^*\bar B$ &  & $-2$ & $\sqrt 2$ \\
\hline
$\Omega_{bb}^*\bar B_s$  &  &  & $-1$\\
\hline
\hline
\end{tabular*}
\end{table*}

\section{Results}
\label{sec:Res}
With the $V_{ij}$ potential of Eq.~\eqref{eq:kernelnr}, we solve the Bethe-Salpeter equation in coupled channels
\begin{equation}\label{eq:BS}
T=[1-VG]^{-1}V,
\end{equation}
where $G$ is the diagonal meson-baryon loop function given by
\begin{align}\label{eq:Gfunc}
  G_l=&i\int\frac{d^4q}{(2\pi)^4}\; \frac{M_l}{E_l(\vec q\,)}\;\frac{1}{k^0+p^0-q^0-E_l(\vec q\,)+i\epsilon}\;\frac{1}{q^2-m_l^2+i\epsilon}\nonumber\\[0.3cm]
     =&\int_{|\vec q\,|<q_{\rm max}}\frac{d^3 q}{(2\pi)^3}\;\frac{1}{2\omega_l(\vec q\,)}
     \;\frac{M_l}{E_l(\vec q\,)}\;\frac{1}{k^0+p^0-\omega_l(\vec q\,)-E_l(\vec q\,)+i\epsilon},
\end{align}
with $p^0$ the energy of the incoming baryon in the meson-baryon rest frame.
$m_l$ and $M_l$ are the meson and baryon masses,
and $\omega_l$ and $E_l$ their energies, $\omega_l=\sqrt{m_l^2+ \vec q^{\;2}}$, $E_l=\sqrt{M_l^2+ \vec q^{\;2}}$.
As in the studies of Refs.~\cite{Xibc66,ViniciusOm,Xibc68},
we use a three-momentum cut-off $q_{\rm max} = 650$ MeV to regularize the loop function.
The poles are searched for on the second Riemann sheet, as done in Refs.~\cite{Xibc66,ViniciusOm,Xibc68},
and the couplings of the states to the different channels are obtained from the residues of the $T_{ij}$ matrix at the pole $z_R$,
knowing that close to the pole one has
\begin{equation}\label{eq:coupling1}
T_{ij}(s)=\frac{g_i\,g_j}{\sqrt s-z_R}.
\end{equation}
The second Riemann sheet is obtained using $G^{II}(s)$ instead of $G(s)$ given by
\begin{equation}\label{eq:RiemSheet2}
G_l^{\uppercase\expandafter{\romannumeral2}}= \left\{\begin{matrix}
G_l(s), & ~~\text{for } \text{Re}(\sqrt{s})<\sqrt{s}_{th,\, l}\\[0.3cm]
G_l(s)+i\displaystyle\frac{2M_lq}{4\pi\sqrt s}, & ~~\text{for } \text{Re}(\sqrt{s})\geq  \sqrt{s}_{th,\, l}
\end{matrix}\right.,
\end{equation}
where $\sqrt{s}_{th,\,l}$ is the threshold mass of the $l$-th channel, and
\begin{equation}\label{eq:momentum}
q=\frac{\lambda^{1/2}(s,m_l^2,M_l^2)}{2\sqrt s}, \quad \text{with}~ {\rm Im}(q)>0.
\end{equation}

In Tables \ref{tab:12} and \ref{tab:13}, we show the couplings and the wave function at the origin for two states with $J^P=\frac{1}{2}^-$
obtained from the coupled channels of Table~\ref{tab:4}.
In addition to the couplings $g_i$,
we show the values of $g_i\,G^{II}_i$ at the pole which according to Ref.~\cite{danijuan} provide the strength of the wave function at the origin.
\begin{table*}[t]
\renewcommand\arraystretch{1.0}
\centering
\caption{\vadjust{\vspace{-0pt}}The $g_i$ couplings of the $10408.18+i93.18$ state (generated dynamically in the PB sector with $J^P=\frac{1}{2}^-$) to various channels and $g_i\,G^{II}_i$ (in MeV).}\label{tab:12}
\begin{tabular*}{0.9\textwidth}{@{\extracolsep{\fill}}ccccc}
\hline
\hline
$\bm{10408.18+i93.18$}& $\Xi_{bb} \,\pi$ & $\Xi_{bb} \,\eta$ & $\Omega_{bb} \,K$ & $\Lambda_b \,\bar B$ \\
\hline
 $g_i$                 &  $1.69+i1.21$  &  $-0.02-i0.09$  &  $-0.86-i0.73$  &  $-1.03-i0.31$\\
 $g_i\,G^{II}_i$  &  $-73.58-i12.87$  &  $0.01+i0.58$  &  $4.52+i5.41$  &  $0.79+i0.39$\\
\hline
                      & $\Sigma_b \,\bar B$ & $\Xi_b \,\bar B_s$ & $\Xi_b^\prime \,\bar B_s$ &\\
\hline
 $g_i$                &  $0.50+i0.23$  &  $-0.28-i0.19$  &  $-0.16-i0.12$&\\
 $g_iG^{II}_i$   &  $-0.29-i0.18$  &  $0.14+i0.12$  &  $0.07+i0.06$&\\
 \hline
\end{tabular*}
\end{table*}

\begin{table*}[h]
\renewcommand\arraystretch{1.0}
\centering
\caption{\vadjust{\vspace{-0pt}}The $g_i$ couplings of the $10686.39+i0.08$ state (generated dynamically in the PB sector with $J^P=\frac{1}{2}^-$) to various channels and $g_i \,G^{II}_i$ (in MeV).}\label{tab:13}
\begin{tabular*}{0.9\textwidth}{@{\extracolsep{\fill}}ccccc}
\hline
\hline
$\bm{10686.39+i0.08}$& $\Xi_{bb}\,\pi$ & $\Xi_{bb}\,\eta$ & $\Omega_{bb} \,K$ & $\Lambda_b \,\bar B$ \\
\hline
 $g_i$                 &$0.01-i0.05$  &  $-0.10+i0.02$  &  $-0.05+i0.04$  &  $0.06+i0.02$\\
 $g_i \,G^{II}_i$& $1.57+i0.51$  & $1.52-i0.23$  & $0.72-i0.52$  & $-0.11-i0.03$\\
\hline
                      & $\Sigma_b \,\bar B$ & $\Xi_b \, \bar B_s$ & $\Xi_b^\prime \, \bar B_s$ &\\
\hline
 $g_i$                & $19.03$  &  $0.02+i0.02$  &$-10.80$&\\
 $g_i\, G^{II}_i$&  $-18.78$  &  $-0.02-i0.02$  & $7.23$&\\
 \hline
\end{tabular*}
\end{table*}

We find two states, one at $10408$ MeV with a width of about $186$ MeV,
which couples mostly to the $\Xi_{bb}\,\pi$ component,
with a non-negligible coupling to $\Omega_{bb} \,K$ and $\Lambda_b \,\bar B$.
The large width of this state stems from the large coupling to the $\Xi_{bb}\,\pi$ channel and the fact that this channel is open.
The second state appears at $10686$ MeV with a very small width.
It couples mostly to the $\Sigma_b \,\bar B$ channel, which is closed.
The $\Xi_{bb}\,\pi$ channel is open, but the coupling to this channel is very small,
which justifies the small width obtained.

Some of the components are quite bound and one may think that these components should be very small.
Yet, this is not the case, since,
as shown in detail in Ref.~\cite{QixinLiang} the size of the channels is not tied to the binding but is determined by the cut-off,
and $r^2 |\psi(r)|^2$ peaks around $r=0.7$\,fm, with still a significant strength around $1$\,fm.

We now consider the states generated from the coupled channels of Table~\ref{tab:6}
from vector-baryon($\frac{1}{2}^+$) states.
We find three states with zero width, degenerate in $J^P=\frac{1}{2}^-, \frac{3}{2}^-$.
We note that the additional pion exchange would break this degeneracy but,
as discussed in Ref.~\cite{QixinLiang}, its effects are largely incorporated in our approach with a suitable choice of $q_{\rm max}$,
and only a small part remains to produce a small splitting between the $\frac{1}{2}^-$ and $\frac{3}{2}^-$ states.
The small difference between the masses of the hidden charm pentaquark states $P_c(4440)$ and $P_c(4452)$ of Ref.~\cite{LHCbpenta},
assumed to be $\frac{1}{2}^-, \frac{3}{2}^-$, respectively, corroborates this finding.

\begin{table*}[t]
\renewcommand\arraystretch{1.0}
\centering
\caption{\vadjust{\vspace{-0pt}}The $g_i$ couplings of the $10732.01+i0$ state (generated dynamically in the VB sector with $J^P=\frac{1}{2}^-, \frac{3}{2}^-$)  to various channels and $g_i \,G^{II}_i$ (in MeV).}\label{tab:14}
\begin{tabular*}{0.6\textwidth}{@{\extracolsep{\fill}}ccccc}
\hline
\hline
$\bm{10732.01+i0}$        &  $\Lambda_b \,\bar B^\ast$ & $\Xi_{bb}\, \rho$     &  $\Xi_{bb} \,\omega$  &  $\Sigma_b \,\bar B^\ast$ \\
\hline
 $g_i$                 &  $-0.01$                   & $0.15$                &  $-0.14$              &  $19.13$  \\
 $g_i\,G^{II}_i$       &  $0.02$                    & $-1.01$               &  $0.88$               &  $-18.72$ \\
\hline
                       &  $\Omega_{bb} \,K^\ast$       & $\Xi_b \,\bar B_s^\ast$ &  $\Xi_{bb}\,\phi$       & $\Xi_b^\prime \,\bar B_s^\ast$  \\
\hline
 $g_i$                 &  $-0.10$                   & $0$                   &  $-0.03$              & $-10.86$  \\
 $g_i \,G^{II}_i$      &  $0.42$                    & $0$                   &  $0.09$               & $7.18$    \\
 \hline
\end{tabular*}
\end{table*}

\begin{table*}[t]
\renewcommand\arraystretch{1.0}
\centering
\caption{\vadjust{\vspace{-0pt}}The $g_i$ couplings of the $10807.41+i0$ state (generated dynamically in the VB sector with $J^P=\frac{1}{2}^-, \frac{3}{2}^-$) to various channels and $g_i \,G^{II}_i$ (in MeV).}\label{tab:15}
\begin{tabular*}{0.6\textwidth}{@{\extracolsep{\fill}}ccccc}
\hline
\hline
$\bm{10807.41+i0}$      &  $\Lambda_b \,\bar B^\ast$    & $\Xi_{bb}\, \rho$         & $\Xi_{bb}\,\omega$    & $\Sigma_b\, \bar B^\ast$   \\
\hline
 $g_i$               &  $7.82$                        & $-0.66$                  & $-0.12$               & $0.06$   \\
 $g_i \,G^{II}_i$    &  $-18.77$                      & $5.52$                   & $0.97$                & $-0.07$ \\
\hline
                     & $\Omega_{bb} \,K^\ast$         & $\Xi_b \,\bar B_s^\ast$  & $\Xi_{bb}\,\phi$      & $\Xi_b^\prime \,\bar B_s^\ast$ \\
\hline
 $g_i$               & $0.16$                         & $7.57$                   & $-0.10$               & $-0.04$  \\
 $g_i\, G^{II}_i$    & $-0.76$                        & $-7.37$                  & $0.39$                & $0.03$  \\
 \hline
\end{tabular*}
\end{table*}

\begin{table*}[h]
\renewcommand\arraystretch{1.0}
\centering
\caption{\vadjust{\vspace{-0pt}} The $g_i$ couplings of the $10869.63+i0$ state (generated dynamically in the VB sector with $J^P=\frac{1}{2}^-, \frac{3}{2}^-$) to various channels and $g_i \,G^{II}_i$ (in MeV).}\label{tab:16}
\begin{tabular*}{0.7\textwidth}{@{\extracolsep{\fill}}ccccc}
\hline
\hline
$\bm{10869.63+i0}$  & $\Lambda_b \,\bar B^\ast$      & $\Xi_{bb} \,\rho$       & $\Xi_{bb} \,\omega$    & $\Sigma_b \,\bar B^\ast$ \\
\hline
 $g_i$           & $0.61$                         & $3.57$                   & $-0.36$               & $-0.37$  \\
 $g_i\,G^{II}_i$ & $-2.25$                        & $-38.50$                 & $3.76$                & $0.53$   \\
\hline
                 & $\Omega_{bb} \,K^\ast$         & $\Xi_b \,\bar B_s^\ast$  & $\Xi_{bb}\,\phi$      & $\Xi_b^\prime \,\bar B_s^\ast$ \\
\hline
 $g_i$           & $-2.41$                        & $1.18$                   & $0.48$                & $0.26$   \\
 $g_i\,G^{II}_i$ & $13.34$                        & $-1.34$                  & $-2.16$               & $-0.21$   \\
 \hline
\end{tabular*}
\end{table*}

In Tables~\ref{tab:14}, \ref{tab:15} and \ref{tab:16}, we show the properties of these three states.
The first state appears at $10732$ MeV and couples mostly to $\Sigma_b \,\bar B^\ast$,
the second is at $10807$ MeV and couples mostly to $\Lambda_b \,\bar B^\ast$,
while the third appears at $10869$ MeV and couples mostly to $\Xi_{bb}\, \rho$.
Note that all channels are closed which is why we obtain zero widths.

Let us look at the states formed from the pseudoscalar-baryon($\frac{3}{2}^+$) channels of Table~\ref{tab:8}.
We find two states, shown in Tables~\ref{tab:17} and \ref{tab:18}.
The first appears at $10447$ MeV with a width of about $186$ MeV.
This state couples mostly to $\Xi_{bb}^* \,\pi$, which is open, justifying the large width.
The second state appears at $10707$ MeV and couples mostly to $\Sigma_b^* \,\bar B$.
The $\Xi_{bb}^* \,\pi$ channel is open, but the small coupling to this channel results in a very small width of this state.

\begin{table*}[t]
\renewcommand\arraystretch{1.0}
\centering
\caption{\vadjust{\vspace{-0pt}}The $g_i$ couplings of the $10447.50+i93.31$ state (generated dynamically in the PB($\frac{3}{2}^+$) sector with $J^P=\frac{3}{2}^-$) to various channels and $g_i\,G^{II}_i$ (in MeV).}\label{tab:17}
\begin{tabular*}{\textwidth}{@{\extracolsep{\fill}}cccccc}
\hline
\hline
$\bm{10447.50+i93.31}$  & $\Xi_{bb}^* \,\pi$    & $\Xi_{bb}^* \,\eta$   & $\Omega_{bb}^* \, K$   & $\Sigma_b^* \,\bar B$  & $\Xi_b^*\,\bar B_s$  \\
\hline
 $g_i$                  & $1.69+i1.21$          & $-0.03-i0.10$         & $-0.87-i0.73$          & $-1.03-i0.49$          & $0.34+i0.26$ \\
 $g_i\, G^{II}_i$       & $-73.61-i12.82$       & $0.05+i0.62$          & $4.58+i5.47$           & $0.60+i0.39$           & $0.15-i0.14$ \\
\hline
\end{tabular*}
\end{table*}

\begin{table*}[t]
\renewcommand\arraystretch{1.0}
\centering
\caption{\vadjust{\vspace{-0pt}}The $g_i$ couplings of the $10706.87+i0.30$ state (generated dynamically in the PB($\frac{3}{2}^+$) sector with $J^P=\frac{3}{2}^-$) to various channels and $g_i\,G^{II}_i$ (in MeV).}\label{tab:18}
\begin{tabular*}{\textwidth}{@{\extracolsep{\fill}}cccccc}
\hline
\hline
$\bm{10706.87+i0.30}$  & $\Xi_{bb}^*\, \pi$   & $\Xi_{bb}^*\, \eta$   & $\Omega_{bb}^* \,K$    & $\Sigma_b^* \,\bar B$    & $\Xi_b^*\, \bar B_s$  \\
\hline
 $g_i$                 & $-0.01+i0.09$        & $0.19-i0.03$          & $0.08-i0.07$           & $19.01-i0.01$            & $-10.79+i0.01$ \\
 $g_i \,G^{II}_i$      & $-3.04-i1.20$        & $-2.55+i0.35$         & $-1.06+i0.91$          & $-18.75-i0.01$           & $7.25+i0.01$  \\
\hline
\end{tabular*}
\end{table*}

Finally, we consider the only ``$\Omega_{bbb}$" state found from the coupled channels of Table~\ref{tab:10}.
This state is at $15212$ MeV and couples mostly to $\Xi_{bb}^*\, \bar B$, as shown in Table~\ref{tab:19}.
All coupled channels are closed and we obtain a zero width for this state.
\begin{table*}[tbph]
\renewcommand\arraystretch{1.0}
\centering
\caption{\vadjust{\vspace{-0pt}}The $g_i$ couplings of the $15212.04+i0$ state (generated dynamically in the PB($\frac{3}{2}^+$) sector with $J^P=\frac{3}{2}^-$ and three $b$ quarks) to various channels and $g_i \,G^{II}_i$ (in MeV).}\label{tab:19}
\begin{tabular*}{0.5\textwidth}{@{\extracolsep{\fill}}cccc}
\hline
\hline
$\bm{15212.04+i0}$    & $\Omega_{bbb}\,\eta$   & $\Xi_{bb}^* \,\bar B$    & $\Omega_{bb}^*\, \bar B_s$  \\
\hline
 $g_i$             & $0.15$                 & $14.03$                  & $-9.82$ \\
 $g_i \,G^{II}_i$  & $-1.44$                & $-18.31$                 & $8.80$ \\
\hline
\end{tabular*}
\end{table*}

In summary we obtained two excited $\Xi_{bb}$ states with $J^P=\frac{1}{2}^-$ coupled to the pseudoscalar-baryon($\frac{1}{2}^+$) channels,
three states with $J^P= \frac{1}{2}^-, \frac{3}{2}^-$, degenerate in our approach,
coupled to the vector-baryon($\frac{1}{2}^+$) channels, two states with $J^P=\frac{3}{2}^-$ coupled to the pseudoscalar-baryon($\frac{3}{2}^+$) channels,
and found only one state corresponding to an excited $\Omega_{bbb}$ state,
coupled to the pseudoscalar-baryon($\frac{3}{2}^+$) channels.

We used the mass of the $\Omega_{bbb}$ ground state from Ref.~\cite{Roberts:2007ni}, $14834$ MeV.
This is quite different from the Lattice QCD calculations in Ref.~\cite{d82}, $14371$ MeV,
similar to Ref.~\cite{Brac}.
Surprisingly, if we redo the calculations using this latter mass,
we obtain a mass of the excited $\Omega_{bbb}$ state of Table \ref{tab:19} which differs by less than $1$ MeV from the former.
The reason is that the obtained excited $\Omega_{bbb}$ state is mostly a $\Xi^*_{bb} \bar B$ molecule and the $\Omega_{bbb}\eta$ channel plays a negligible role.
This is due to the zero $D_{ii}$ coefficient for $\Omega_{bbb}\eta$ in Table \ref{tab:11},
which indicates that there is no direct interaction in this channel.
The negligible effect of this channel in the excited $\Omega_{bbb}$ state can also be seen in the small coupling to this channel, $0.15$ versus $14.03$ for the coupling to the $\Xi^*_{bb} \bar B$ channel.
The latter channel is bound by about $300$ MeV,
which again is due to the scale of the masses and the large $D_{ii}=-2$ coefficient for the diagonal $\Xi^*_{bb} \bar B \to \Xi^*_{bb} \bar B$ transition.

\section{Conclusions}
We carried out a study of the interactions of meson-baryon coupled channels that lead to the formation of bound or resonant states,
corresponding to the excited $\Xi_{bb}$ and $\Omega_{bbb}$ states.
As in related studies of $\Xi_c, \Xi_b, \Xi_{bc}$ and hidden charm molecular states,
we used an interaction based on the exchange of vector mesons,
which in the case of light quarks gives rise to the chiral Lagrangians.
In particular, the exchange of light vectors, which produces the dominant part of the interaction,
leaves the heavy quarks as spectators and fulfills the rules of heavy quark symmetry.
We find seven $\Xi_{bb}$ states and one $\Omega_{bbb}$ state of molecular nature.
The success in describing the hidden charm pentaquark states,
and of some $\Omega_c, \Xi_c, \Xi_b$ states using the same input for the interaction,
supports our confidence that the predictions are realistic. It will be interesting to compare them with the future measurements which are likely to be made by LHCb and Belle II.

\section{ACKNOWLEDGEMENT}
We thank the hospitality of Guangxi Normal University, where the main part of this work was done.
This work is partly supported by the National Natural Science Foundation of China under Grants No. 11975083, No. 11947413,
No. 11847317, No. 11565007, No. 11735003 and No. 1191101015.
Q.X Yu acknowledges the support from the National Natural Science Foundation of China
(Grants No. 11775024, No. 11575023 and No. 11805153) and China Scholarship Council.
This work is partly supported by the Spanish Ministerio de
Economia y Competitividad and European FEDER funds under Contracts No. FIS2017-
84038-C2-1-P B and No. FIS2017-84038-C2-2-P B, and the Generalitat Valenciana in the
program Prometeo II-2014/068, and the project Severo Ochoa of IFIC, SEV-2014-0398.


\end{document}